# The generation of warm dense matter using a magnetic anvil cell

P.-A. Gourdain[1,2], A. B. Sefkow[3,2], C. E. Seyler[4]

[1] *Extreme State Physics Laboratory, Physics and Astronomy Department, University of Rochester, Rochester NY 14627*
[2] *Laboratory for Laser Energetics, University of Rochester, Rochester NY 14627*
[3] *Mechanical Engineering Department, University of Rochester, Rochester NY 14627*
[4]*Laboratory for Plasma Studies, Cornell University, Ithaca, NY 14850*

*Abstract*—**Warm dense matter is present at the heart of gas giants and large exo-planets. Yet, its most basic properties are unknown and limit our understanding of planetary formation and evolution. In this state, where pressure climbs above 1 Mbar, matter is strongly coupled and quantum degenerate. This combination invalidates most theories capable of predicting the equation of state, the viscosity or heat conductivity of the material. When such properties are missing, understanding planetary evolution becomes an arduous endeavor. Henceforth, research in this field is actively growing, using high power laser or heavy ion beams to produce samples dense enough to overcome the 1 Mbar limit. These samples are not actively confined and tend to expand rapidly, precluding the existence of any thermodynamically stable equilibrium. However, a mega-ampere-class pulsed-power generator can produce confined matter in the Mbar range, providing two conditions are being met. First, the sample needs to be compressed cylindrically, to maximize magnetic pressure (compared to slab compression). Second, a damper must be used to preclude the formation of a corona around the sample. This corona robs the main sample from valuable current and limits the homogeneity of the compression. According to numerical simulations, the setup proposed here, and called a magnetic anvil cell, can reach pressure on the order of 1 Mbar using a mega-ampere pulsed power driver. These samples span several millimeters in length. Unlike diamond anvil cell, which pressure is limited below 1 Mbar due to materials strength, the magnetic anvil cell has virtually no pressure limit. Further, the current heats the sample to several eV, a temperature well beyond diamond anvil cell capabilities.**

*Index Terms*—**Warm dense matter, Z-pinch, pulsed-power generator, diamond anvil cell**

## I. INTRODUCTION

THIS paper presents numerical simulations that demonstrate how pulsed-power drivers can generate large samples of warm dense matter, using a setup called a magnetic anvil cell. By "large" we mean on the order of the material grain size, or larger, i.e. mesoscale. Warm dense matter (WDM) can be found in the core of giant planets, at the surface of brown dwarves or at early times in implosions for inertial fusion experiments. It is a state of matter where ions are strongly coupled and electrons are partially degenerate. While being simultaneously in a plasma state and a condensed matter state, both plasma and condensed matter theories are unable to capture its basic properties. Going back to first principles and using density functional theory would prove impractical to model macroscopic systems. When dealing with planetary formation or fusion experiments, macroscopic conservation laws (e.g. MHD) are the only practical approach to numerical simulations. This path demands accurate models for the equation of state and transport coefficients like viscosity, heat or electrical conductivities. To benchmark theories capable of capturing the microscopic properties of warm dense matter at the macroscopic scale, experimentalists need to generate large (>1 mm) samples of warm dense matter. At this size, the velocity shear or temperature gradients can be measured effectively using a bright x-ray light source. While pulsed-power drivers where first used to study warm dense matter, they

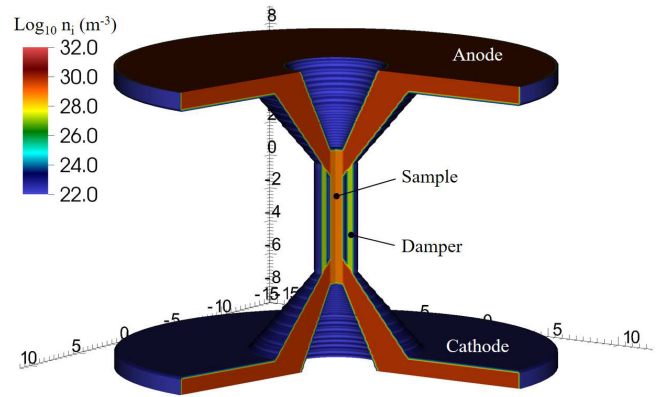

Fig. 1. A three-dimensional view of the magnetic anvil cell. All distances are in mm. The anode and cathode have the density of copper-tungsten. The sample has the density of aluminum. The damper has variable density, given in the text.

This research was supported by the DOE grant number DE-SC0016252.

The next few paragraphs should contain the authors' current affiliations, including current address and e-mail. For example, F. A. Author is with the National Institute of Standards and Technology, Boulder, CO 80305 USA (e-mail: author@ boulder.nist.gov).

S. B. Author, Jr., was with Rice University, Houston, TX 77005 USA. He is now with the Department of Physics, Colorado State University, Fort Collins, CO 80523 USA (e-mail: author@lamar.colostate.edu).

T. C. Author is with the Electrical Engineering Department, University of Colorado, Boulder, CO 80309 USA, on leave from the National Research Institute for Metals, Tsukuba, Japan (e-mail: author@nrim.go.jp).

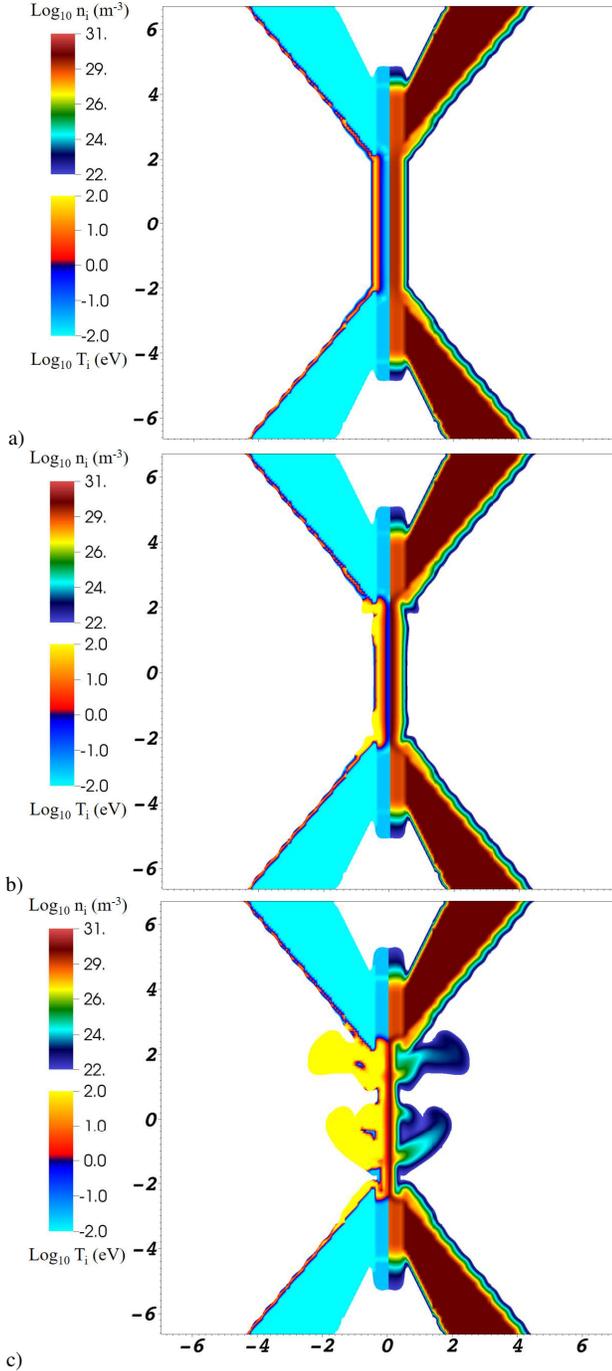

a)

b)

c)

Fig. 2. Time evolution of the MAC for a) t=140 ns, b) 160 ns and c) 180 ns. Each panel shows the ion density (on the right) and the ion temperature (on the left), plotted on the logarithmic scale. All dimensions are in mm. the regions were the ion density is smaller than $10^{22}$ m$^{-3}$ are not plotted.

to compress matter with a pulsed-power driver. As in the diamond anvil cell, a material damper is necessary to compress the central sample. The damper isolates the sample from external current instabilities. At 1 MA, pressures as large as 2 Mbar can be obtained. Unlike diamond anvil cell, which pressure is limited by materials strength, the magnetic anvil cell has virtually no pressure limit. The damper material can be a metal, an insulator or a pre-ionized gas. We study here how the density and resistivity of the damper impacts the quality of the implosion and the pressure of the central sample. After this introduction, the paper lists the basic elements of the MAC. Then two dimensional numerical simulations show the dynamics expected in the MAC. Finally, we look at the impact of damper mass density and electrical resistivity on the properties and the quality of the final WDM sample.

## II. INITIAL CONDITIONS

The experimental setup is shown in Fig. 1. The MAC is composed of 2 cylinders. The innermost cylinder will be compressed by the pulsed-power generator and will turn into a warm dense matter state as the current of the generator increases. The outer cylinder is the damper. The top and bottom electrodes are conical, allowing a smooth transition of current from the generator connections down to the much smaller load.

The damper fulfils three different functions. First, the damper precludes the formation of a corona (i.e. ablated plasma) around the central sample. This corona diverts part of the current away from the main sample. As the current flows inside the corona, the corona temperature increases and its resistivity decreases, leading to larger currents inside the corona. Ultimately, part of the energy delivered by the driver is used to heat the corona and not to compress the sample directly. Second, the damper is limiting the formation [1] of electrothermal instabilities [2] at the surface of the central sample. Finally, the density and the resistivity of the damper can be chosen to optimize the compression for a given current. For identical samples, a heavy damper will slow down the implosion and reduce the overall pressure on the central sample at peak current while a lighter damper will allow for higher pressures. If the damper resistivity is large, then most of the current flows inside the sample and generate a coronal plasma which is difficult to compress, leading to smaller stagnation pressures. Since large current densities can also generate electrothermal instabilities[3], the presence of the damper slow down their growth.

This paper looks at the impact of the density and resistivity of the damper on the pressure and overall quality of the implosion of the sample. We used the extended-MHD code PERSEUS [4] in cylindrical coordinates to simulate the MAC on a mega-ampere class pulsed-power generator like HADES [5]. HADES is a 250GW pulsed power driver using linear transformer technology [6] to generate a current of 1 MA in less than 150 ns. The time evolution of the current is given by

$$I = I_{peak} sin\left(\frac{\pi}{2}\frac{t}{t_r}\right) exp\left(-\frac{t}{t_d}\right) \quad (1)$$

where $I_{peak}$ is the peak current, $t_r$ is the current rise time and $t_d$ is fell out of favor due to the excessive corona building a round the sample and precluding homogeneous implosion. Experiments using kilojoule lasers and heavy ion beams have been successful so far, but have some drawbacks: small sample size and bulk matter not completely at equilibrium. The sample size of limited by the amount of available energy. The lack of confinement precludes materials studies on time scale long enough to allow for proper equilibrium (e.g. strain redistribution). The magnetic anvil cell (MAC) was designed to mitigate both limitations. This configuration uses a z-pinch the current damping time [7]. While simulations have been performed in two dimensions in this paper, Reference [8] showed that three-dimensional instabilities are not detrimental to the overall compression of the sample. So, the present work focuses on the impact damper properties on sample compression ignoring three-dimensional instabilities. Clearly, these instabilities must be taken into account to truly define the parameter space that the MAC can cover for a 1 MA driver.

The simulation domain covers 15 mm horizontally and 18 mm vertically. The total number of cells is on the order of 60,000. The cell size is 70 μm. This resolution does not allow to estimate precisely the maximum pressure that can be reached in the proposed setup since the code cannot compress the rod beyond 70 μm. At stagnation, the maximum pressure can be computed from the equilibrium equation

$$\vec{\nabla} p = \vec{J} \times \vec{B} \ . \tag{1}$$

This equation gives a maximum pressure of 10 Mbar using 70

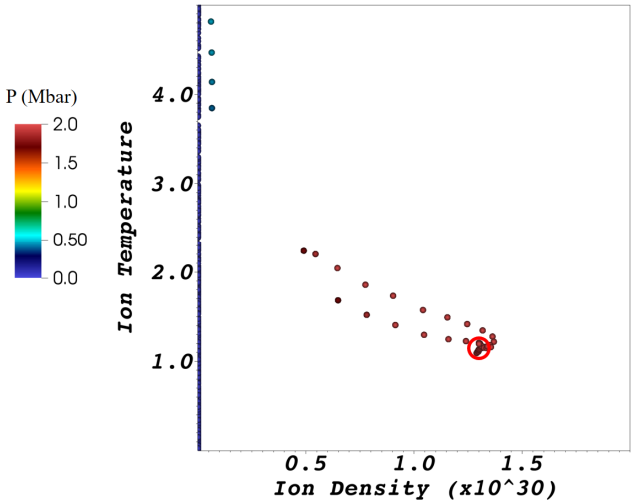

Fig. 4. Ion temperature versus ion density on the logarithmic scale at t=180ns for a collection hockey puck encompassing 2 mm above and below the mid-plane, for a foam density 30 times smaller than solid density. When the collection volume is limited to 1 mm above and below the midplane we obtain the data points located inside the red circle. The ion density is given in m$^{-3}$ and the ion temperature is in eV.

μm as the smallest allow radius. So, it is reasonable to infer that any pressure below 10 Mbar is realistically computed by the code.

### III. IMPLOSION DYNAMICS OF THE MAGNETIC ANVIL CELL

First, we look at the implosion dynamics of the cell and show how it compares to the gas anvil cell previously published [8]. The simulation spanned 300 ns, with a current rise time $t_r$ of 150 ns and a damping time of $6t_r$. In the case presented here, the damper is made from a foam, 30 times lower than solid density.

Fig. 2 shows the simulation results for the central load region near current peak. Each panel is split into two plots. On the right side, the panel shows the ion density on the logarithmic scale. The left side gives the ion temperature, also on the logarithmic scale. At this stage, the damper has fully merged with the sample. The stagnation happens at 185 ns. At this time, the pressure is on the order of 2 Mbar, below the 10 Mbar limit discussed in the previous section. As a result, the maximum compression is reached due to driver (i.e. energy) limitation rather than the intrinsic limitation of the simulation resolution. By the time the implosion reaches stagnation, density spikes have formed around the dense column. The simulation shows three distinct locations where low-density spikes protrude outwards, indicating potential inhomogeneities in the compression of the sample.

Reducing the inhomogeneities caused by local instabilities is primordial to the success of the MAC. Radial inhomogeneities are perfectly acceptable if axisymmetry is enforced. Abel inversion or LIDAR measurement techniques can be performed under these conditions. So, even if the resolution of the simulation cannot resolve below 70 μm, any inhomogeneity along this direction does not pose a major problem. However, if viscosity measurements are to be successful, axial homogeneities are more problematic, since it is along this direction that the measurement will be most meaningful. We can quantitatively assess the impact of inhomogeneities on the pressure by plotting the ion temperature as a function of the ion density inside a chosen volume of the simulation domain. Fig. 4 shows the result of two collection volumes. When the collection volume is a hockey puck of 15 mm radius, spanning a height of 2 mm, with 1 mm above and 1 mm below the midplane, all the data points falls into a red circle. In this case, the variation in density is less than 10$^{29}$ m$^{-3}$ and 0.1 eV (or 10% for both), quite an acceptable variation. However, when we extend the collection volume to encompass the full actively compressed sample, i.e. 2 mm above and below the mid plane, the data spreads well outside of the red circle. There is an obvious dispersion of data points in this case, well above 100%. This simulation shows that only the central 2 mm of the sample can be used effectively. The maximum pressure is reached closest to the mid-plane, an indication that parasitic behavior associated with end effects will have limited impact on the central region of the sample.

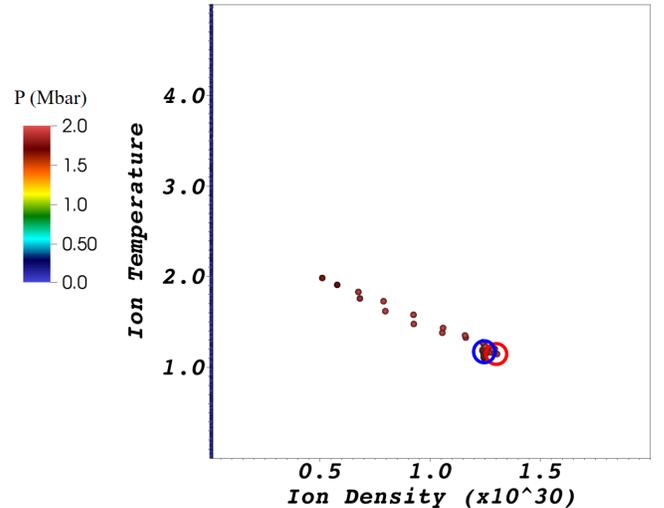

Fig. 3. Ion temperature versus ion density on the logarithmic scale at t=180ns for a collection hockey puck encompassing 2 mm above and below the mid-plane, for a foam density 100 times smaller than solid density. When the collection volume is limited to 1 mm above and below the midplane we obtain the data points located inside the blue circle. The red circle from Fig. 4 is shown here for comparison. The ion density is given in m$^{-3}$ and the ion temperature is in eV.

Overall, we have found that this axial dependence is relatively well bounded in the MAC proposed here. While higher resolutions or three-dimensional simulations may reveal inhomogeneities at smaller scales, the simulation presented here shows encouraging behavior at the 100 µm scale.

## IV. IMPACT OF THE FOAM DENSITY

We now look at the implosion dynamics of the cell when the foam density is 3 times lower. All the other parameters are kept constant. In this case, we arrive to stagnation in 160 ns, 20 ns earlier than in the previous case. Fig. 3 shows that there is only a minor difference between the two cases. The data point spread and overall dynamics are very similar. This indicates that the mass of the damper weakly impacts the properties of the final sample. Regardless, a damper with much lower density would be difficult to manufacture and set in place experimentally. A much denser damper would start to interfere with the x-rays used to probe the properties of the sample in the warm dense matter state. As a result, the foam density found here is well optimized.

## V. IMPACT OF THE FOAM RESISTIVITY

After looking at the impact of the foam density, we turn to its resistivity. In the initial case shown in previous figures, the foam had a resistivity following a version of the Lee-More [9] and Desjarlais [10] resistivity specifically tailored for aluminum. We now artificially increased this resistivity by a factor of 10, 100 or 1000, compared to this reference resistivity $\rho_{ref}$. This artificial increase takes place inside the damper only.

Fig. 5 shows the impact of the damper resistivity on the sample pressure at stagnation. As before, end effects start to be noticeable 1 mm above and below the mid plane. However, they have a minor impact on the central region of the sample. What is more remarkable is the impact of the damper resistivity on the pressure itself. Paradoxically, as the resistivity increases and more current starts to flow inside the sample, the sample pressure diminishes at stagnation.

A similar effect was seen when a pre-ionized gas puff was used [11]. When the current flows inside the sample, the formation of a corona is unavoidable early in the discharge, when the magnetic field is low. This corona heats up. It then partially shields the sample from the external magnetic field. The pressure also increases rapidly, and the plasma goes from high beta to low beta.

In the presence of a pre-ionized gas puff, the current hardly flows inside the central sample at the beginning of the discharge. It only flows later, when the gas puff has fully imploded onto the sample and when the magnetic field surrounding the sample is already large. This large magnetic field prevents the ablated plasma from expanding and heating up. In this respect, the gas performs the function of a plasma opening switch.

A similar effect happens here. When the damper resistivity is high, the current flows mostly inside the sample, early in the discharge. So, the formation of the corona happens under a weak magnetic field that cannot quench the formation and expansion of the corona. In fact, the remnants of the corona are still present at stagnation, visible to the right of the vertical axis in Fig. 5-b and c. When the damper resistivity is low, a sizable portion of the current flows inside the damper, limiting the formation of the corona, which has now disappeared from Fig. 3, Fig. 4, and Fig. 5-a. The full current is forced back into the sample only when the damper has fully imploded onto the sample, ramming both corona and magnetic field into the sample. This process is more efficient and the effective pressure is higher, as shown in Fig. 5-b and c.

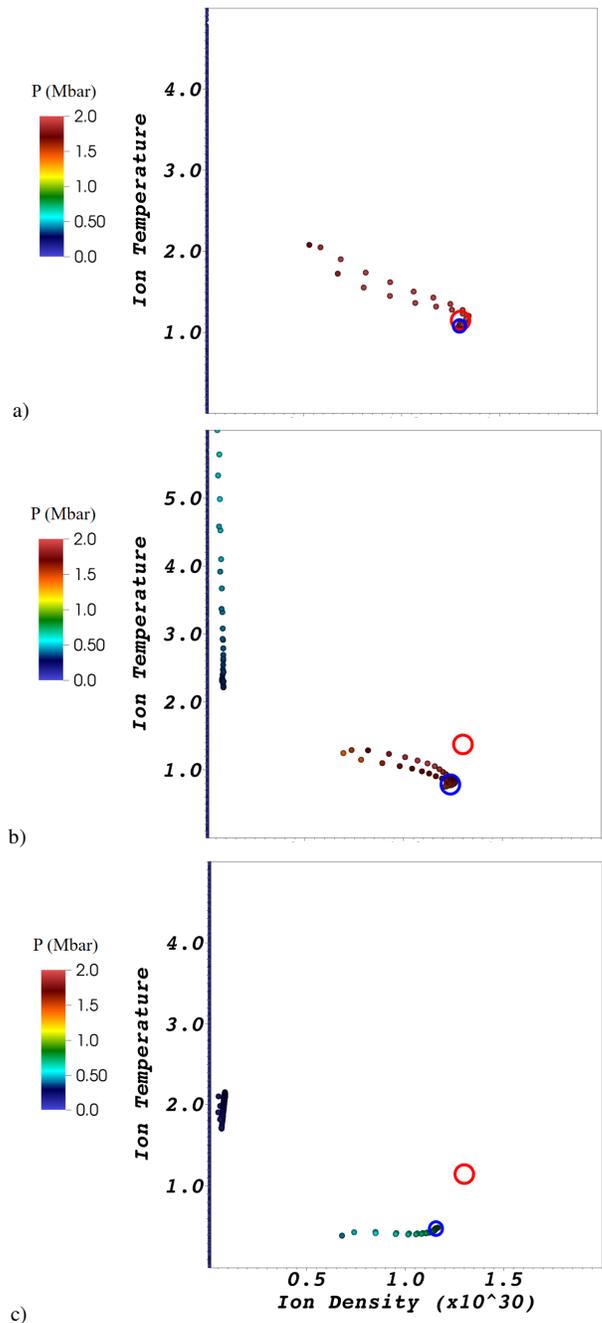

Fig. 5. Ion temperature versus ion density on the logarithmic scale at t=180ns for a collection hockey puck encompassing 2 mm above and below the mid-plane, for a foam resistivity a) 10 times, b) 100 times and c) 100 lower than the initial foam resistivity. When the collection volume is limited to 1 mm above and below the midplane we obtain the data points inside the blue circle. The red circle from Fig. 4 is shown here for comparison. The ion density is given in m^-3 and the ion temperature is in eV.

When the resistivity is kept low ($\rho{\sim}10\rho_{\text{ref}}$), the stagnation pressure of the sample stays around 2 Mbar, mostly unchanged compared to the reference case. As the resistivity increases ($\rho{\sim}100\rho_{\text{ref}}$ then $\rho{\sim}1000\rho_{\text{ref}}$) there is more and more current flowing inside the sample and the pressure at stagnation is smaller, as just explained. This effect can give a supplementary degree of freedom, allowing to produce lower pressures using the same discharge current. The change in resistivity does not impact the quality of this implosion in the central region, above and below the midplane.

## VI. Conclusion

This paper showed how a foam damper is an effective means to control and improve the compression of large scale cylindrical samples. The effect of the foam density is relatively weak on stagnation pressure. A change in foam resistivity impacts more dramatically the stagnation pressure. During the implosion, all the way to stagnation, the central region of the sample implodes relatively homogeneously and the pressure reaches 2 Mbar along 2 mm of the sample axis. The variations in ion density and temperature are found to be smaller than +/- 10%. While further studies are required to assess the quality of the sample under megabar pressures, the MAC shows great promise in the study of materials under extreme pressures at the mesoscale.